\documentclass[10pt,twocolumn,english]{revtex4-1}
\usepackage[T1]{fontenc}
\usepackage[latin9]{inputenc}
\usepackage[active]{srcltx}
\usepackage{amsmath}
\usepackage{amssymb}
\usepackage{graphicx}

\makeatletter
\usepackage{babel}

\makeatother

\usepackage{babel}
\begin{document}

\title{Embedding the dynamics of a single delay system into a feed-forward
ring }

\author{{\normalsize{}Vladimir~Klinshov$^{1}$, Dmitry Shchapin$^{1}$,
Serhiy~Yanchuk$^{2}$, Matthias Wolfrum$^{3}$, Otti D'Huys$^{4}$,
and Vladimir Nekorkin$^{1}$}}

\affiliation{$^{1}$Institute of Applied Physics of the Russian Academy of Sciences,
46 Ul'yanov Street, 603950, Nizhny Novgorod, Russia}

\affiliation{$^{2}$Technical University of Berlin, Institute of Mathematics,
Straße des 17. Juni 136, 10623 Berlin, Germany}

\affiliation{$^{3}$Weierstrass Institute for Applied Analysis and Stochastics,
Mohrenstr. 39, 10117, Berlin, Germany }

\affiliation{$^{4}$Aston University, Department of Mathematics, B4 7ET Birmingham,
United Kingdom}
\begin{abstract}
We investigate the relation between the dynamics of a single oscillator
with delayed self-feedback and a feed-forward ring of such oscillators,
where each unit is coupled to its next neighbor in the same way as
in the self-feedback case. We show that periodic solutions of the
delayed oscillator give rise to families of rotating waves with different
wave numbers in the corresponding ring. In particular, if for the
single oscillator the periodic solution is resonant to the delay,
it can be embedded into a ring with instantaneous couplings. We discover
several cases where stability of periodic solution for the single
unit can be related to the stability of the corresponding rotating
wave in the ring. As a specific example we demonstrate how the complex
bifurcation scenario of simultaneously emerging multi-jittering solutions
can be transferred from a single oscillator with delayed pulse feedback
to multi-jittering rotating waves in a sufficiently large ring of
oscillators with instantaneous pulse coupling. Finally, we present
an experimental realization of this dynamical phenomenon in a system
of coupled electronic circuits of FitzHugh-Nagumo type.
\end{abstract}
\maketitle

\section*{Introduction}

A specific feature of systems with time delayed feedback is that their
phase space is infinite dimensional. In particular, for large delay
times, this can lead to complex high dimensional dynamics \cite{Lang1980,Giacomelli1998,Wolfrum2006,Wolfrum2010,Larger2012,Larger2013,Soriano2013,YanchukGiacomelli2017}.
Similarly, complex high dimensional dynamics can be observed in large
network systems \cite{Kuramoto2002,Pikovsky2001,RodriguesPeronJiEtAl2016,Boccaletti2006}.
In this case, the presence of long feed-forward loops can play a similar
role to a direct feedback with a large delay time. 

In this paper, we will substantiate this heuristic correspondence
and provide a detailed study of the relation between periodic orbits
of oscillators with time delay and rotating waves in corresponding
systems where many oscillators of this type are coupled in a feed-forward
ring structure. The dynamics of such rings have been studied before,
including rotating waves and patterns \cite{Bonnin2009,Perlikowski2010,Bungay2007},
chaotic and complex dynamics \cite{Marino1999,Zhang2001,Dhuys2012}
and transient processes \cite{Horikawa2009a,horikawa2012b}. The influence
of coupling delays has been also investigated theoretically \cite{Dodla2004,Kantner2013,Sande2008,DHuys2008}
and experimentally \cite{Sanchez1998,Williams2013,Woafo2004,Klinshov_2016,Rosin13,Dhuys16Chaos,Lohman2017}. 

We start from the basic observation that a periodic solution of a
single oscillator with delayed feedback can be embedded into the feed-forward
ring if certain relations between the delay times of the feedback
oscillator and the delay in the coupled ring are satisfied. For periodic
solutions with a period that is resonant, i.e. relates rationally
to the delay time, the embedding is possible in a ring with instantaneous
coupling, i.e. without delay. Furthermore, we discuss several cases
where the stability properties of the periodic solution of the feedback
oscillator and the rotating wave in the ring are related.

An important tool for this investigation is the multi-scale approach
to systems with large delay \cite{YanchukGiacomelli2017,Lichtner2011,Sieber2013},
where the singular limit of the delay time tending to infinity is
used to obtain an asymptotic description of the stability. Note that
a periodic orbit of a delay differential equation is again a solution
when the delay is increased by an integer multiple of the period \cite{Yanchuk2009}.
In this way the orbit reappears for a sequence of delay times and
can be found for arbitrary large delays. A similar approach can be
used not only in the case when both the feedback delay of the single
oscillator and the coupling delay in the ring are large, but also
in the case of instantaneous coupling in a sufficiently large ring.
Another similarity between delay systems and rings was mentioned in
\cite{Perlikowski2010a,Yanchuk2008a,Perlikowski2010}, where certain
important features of the spectrum of steady states and periodic solutions
in systems with large delay and rotating waves in large feed-forward
rings are reported. In Refs.~\cite{Dhuys2012,VanderSande2008}, it
is shown that the correlation properties of a ring with $N$ elements
can be deduced from the autocorrelation of the single delayed feedback
system.

A specific example for the emergence of complex dynamics in systems
with delay is the recently discovered regime of multi-jittering in
oscillators with delayed pulse feedback \cite{Klinshov2015PRL,Klinshov2015a,Klinshov_2016,Klinshov2016}.
Based on the theoretical background given in the first part of the
paper, we show that this phenomenon has its counterpart in jittering
rotating waves in a feed-forward pulse-coupled ring. Their appearance
can even be demonstrated experimentally by a system of coupled electronic
FitzHugh-Nagumo circuits.

The structure of the paper is as follows. Section \ref{sec:Periodic-solutions-of}
presents theoretical results on the correspondence between periodic
solutions of the delayed oscillator and the ring. It also discusses
the cases when the stability of corresponding periodic solutions is
related. Further, Sec.~\ref{sec:Instantaneous-coupling-in} explains
how periodic solutions of a delay feedback oscillator are embedded
as rotating waves in a ring with instantaneous coupling under the
condition of resonance between period and delay. In Sec.~\ref{sec:Jittering-regimes-in}
we show how the multi-jittering bifurcation known from delayed systems,
appears in feed-forward rings with instantaneous couplings. In particular,
in Sec.~\ref{subsec:Experimental-demonstration-in}, an experimental
demonstration of the jittering waves is presented in a ring of electronic
FitzHugh-Nagumo oscillators. 

\section{Periodic solutions of a single delayed oscillator and rotating waves
in a unidirectionally coupled ring \label{sec:Periodic-solutions-of}}

We start with a single oscillator with delayed feedback 
\begin{equation}
\frac{dx(t)}{dt}=f\left(x(t),x(t-\tau)\right),\label{eq:1}
\end{equation}
where $x\in\mathbb{R}^{d}$, and $\tau\geq0$ is the time-delay. We
assume that for $\tau=\tau_{0}$ equation (\ref{eq:1}) has a periodic
solution $h(t)$ with a period $T$. It is easy to see that changing
the delay, $h(t)$ reappears as a periodic solution whenever multiple
of its period $T$ is added to the delay, that is, $h(t)$ is a periodic
solution for all values of the delay 
\begin{equation}
\tau_{k}=\tau_{0}+kT,\qquad k\in\mathbb{N}.\label{eq:reappearance}
\end{equation}
Note that without loss of generality, we can assume $0\leq\tau_{0}<T$,
i.e. $\tau_{0}$ is the minimal positive delay at which the given
solution $h(t)$ exists, such that for $k\geq0$ we obtain all non-negative
values of the delay for which the periodic solution $h(t)$ reappears.
For a detailed exposition of this reappearance phenomenon and its
consequences see \cite{Perlikowski2010}.

Now we take $N$ oscillators of the form (\ref{eq:1}), where, instead
of the delayed self-feedback, we employ a delayed coupling to the
neighboring oscillator: 
\begin{equation}
\frac{dx_{n}(t)}{dt}=f(x_{n}(t),x_{n-1}(t-\sigma))\label{eq:ring}
\end{equation}
Considering the oscillator index $n=1,...,N$ modulo $N$, we obtain
a unidirectionally coupled ring, where we denote the coupling delay
by $\sigma$ in order to distinguish it from the feedback delay $\tau$
in the single oscillator (\ref{eq:1}). Following Refs. \cite{DHuys2014,DHuys2016}
the periodic solution $h(t)$ of the single oscillator can be used
to construct rotating waves for the ring system (\ref{eq:ring}) by
\begin{equation}
x_{n}(t)=h(t+n\theta),\label{eq:rotwave}
\end{equation}
where $\theta$ is the phase lag between neighboring oscillators.
In order to satisfy the periodic boundary condition in the ring, this
phase lag has to be resonant to the period $T$, i.e. 
\[
\theta=\theta_{M}=MT/N
\]
where $M=0,...,N-1$ is the wave number in the ring. It can be easily
seen that the ansatz (\ref{eq:rotwave}) leads to a rotating wave
solution of (\ref{eq:ring}), if the delays in the two systems satisfy
the compatibility condition 
\begin{equation}
\sigma=\sigma_{k,M}=\tau_{0}+kT-\theta_{M}=\tau_{k}-\theta_{M},\qquad k\in\mathbb{N}.\label{eq:delaycomp}
\end{equation}
In this way, we obtain for each fixed wave number an increasing sequence
of delay times $\sigma_{k,M}$ for which the periodic solution $h(t)$
of the single oscillator with delayed feedback (\ref{eq:1}) can be
embedded as a rotating wave into the unidirectionally coupled ring
system (\ref{eq:ring}). Note that for some choices of $M$ the first
value $\sigma_{0,M}$ can be negative and the first positive delay
value in the reappearance sequence is $\sigma_{1,M}$.

\begin{figure}
\includegraphics{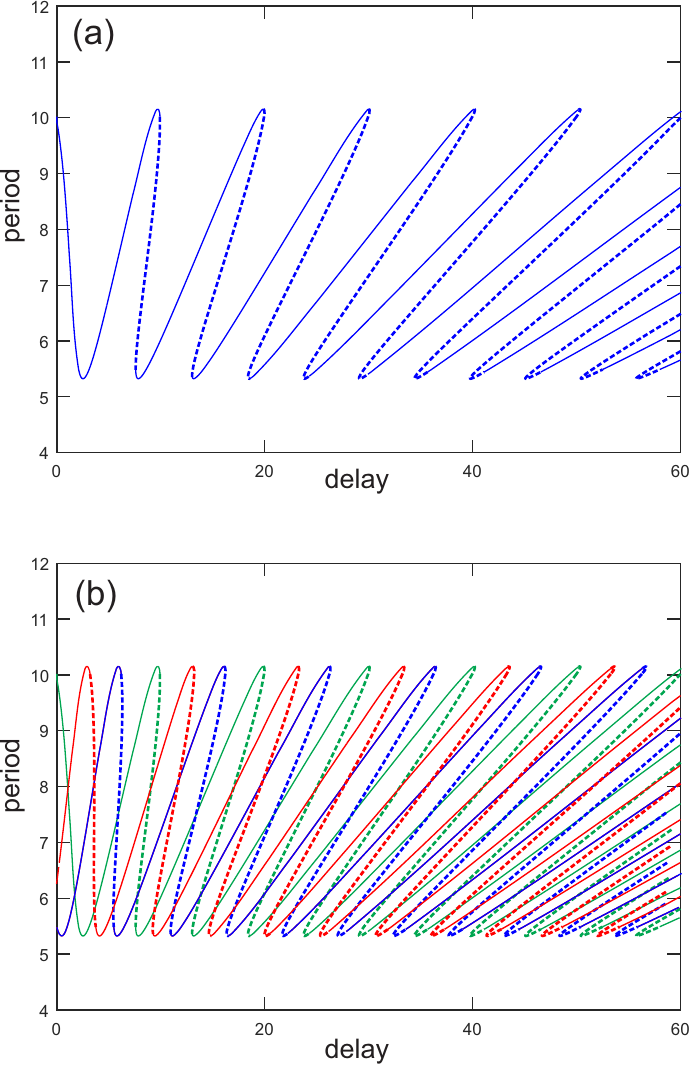}

\caption{\label{fig:vdp_bd}(a) Branch of periodic solutions for a single Van-der-Pol
oscillator with delayed feedback (\ref{eq:VdP}). Solid/dashed parts
indicate stability/instability. (b) Branches of rotating waves for
a feed-forward ring of $N=3$ Van-der-Pol oscillators with delayed
coupling (\ref{eq:VdP}). Colored parts indicate stability for $M=0$
(blue), $M=1$ (red), and $M=2$ (green).}
\end{figure}

The described relation between a single oscillator and a feed-forward
ring is illustrated in Fig. \ref{fig:vdp_bd}. In Fig.~\ref{fig:vdp_bd}(a)
we show the branch of periodic solutions for the Van-der-Pol oscillator
with delayed feedback 
\begin{equation}
\frac{d^{2}x(t)}{dt^{2}}=\alpha\left(1-x(t)^{2}\right)\frac{dx(t)}{dt}-x(t)+\kappa x(t-\tau),\label{eq:VdP}
\end{equation}
plotting the period $T$ versus the delay time $\tau$. Such form
of the bifurcation diagram is typical when the period on the branch
is bounded (see more details in \cite{Yanchuk2009}) and it clearly
leads to an increasing coexistence of periodic solutions with of time-delay.
Note the increasing skewness of the branch that can be explained by
the period depending spacing of the reappearance sequences. 

For the ring of $N$ oscillators with unidirectional coupling 
\begin{equation}
\frac{d^{2}x_{n}(t)}{dt^{2}}=\alpha\left(1-x_{n}(t)^{2}\right)\frac{dx_{n}(t)}{dt}-x_{n}+\kappa x_{n-1}(t-\sigma),\label{eq:VdPring}
\end{equation}
in the case of $N=3$ we obtain three branches of rotating waves shown
in Fig.~\ref{fig:vdp_bd}(b). While the branch for wave number $M=0$
coincides with the periodic solution for the single oscillator, shown
in panel (a), the other two branches can be obtained by a shift of
$\theta_{M}=MT/N$ along the horizontal axis. The stability regions
are highlighted as solid parts of the branch, unstable as dashed.
The stability regions vary from branch to branch, however, they turn
out to coincide for large delays. This phenomenon is explained analytically
in the following section.

\subsection*{Stability properties}

\label{sec:stab_prop} A reappearing periodic solution $h(t)$ of
the delay equation (\ref{eq:1}) has typically different stability
properties for different values of $k$ in the reappearance sequence
(\ref{eq:reappearance}). For large values of $k$, corresponding
to large values of the delay $\tau_{k}$, the stability can be asymptotically
described using the singular limit $k\rightarrow\infty$. The theoretical
background has been elaborated in detail in \cite{Yanchuk2009,Sieber2013}.
It is shown there that in this case the Floquet-spectrum consists
of pseudo-continuous spectrum, converging to continuous curves, and
strongly unstable spectrum, converging to finitely many isolated points.
These limiting objects can be determined from a multi-scale approach
and then be used as a criterion for stability or instability for finite
but sufficiently large delays $\tau_{k}$.

Based on this approach, we will show now that the periodic solution
$h(t)$ of the single oscillator (\ref{eq:1}) and the corresponding
rotating wave solutions (\ref{eq:rotwave}) of the ring system (\ref{eq:ring})
actually lead to the same limiting spectra and hence share the same
linear stability properties, if in both cases the delay is sufficiently
large, i.e. $k\rightarrow\infty$ (cf.~\cite{DHuys2014,DHuys2016}).

The linear stability of the solution $h(t)$ in (\ref{eq:1}) is given
by the following variational equation for a perturbation $\delta(t)\in\mathbb{R}^{d}$:
\[
\frac{d\delta(t)}{dt}=A(t)\delta(t)+B(t)\delta(t-\tau),
\]
where $A(t)=\partial_{1}f(h(t),h(t-\tau))$ and $B(t)=\partial_{2}f(h(t),h(t-\tau))$
are the derivatives of $f$ with respect to its first and second argument,
respectively. To obtain the corresponding Floquet problem, we substitute
$\delta(t)=p(t)e^{\lambda t}$ and get 
\begin{equation}
\frac{dp(t)}{dt}=(A(t)-\lambda\mathrm{Id})p(t)+e^{-\lambda\tau}B(t)p(t-\tau),\label{eq:eig1}
\end{equation}
The set of complex values $\lambda$ for which Eq.~(\ref{eq:eig1})
admits a $T$-periodic solution $p(t)$ are the characteristic exponents.
Note that the coefficient matrices $A(t)$ and $B(t)$ are $T$-periodic
as well. The solution $h(t)$ is stable if all characteristic exponents
have negative real parts.

Similarly, the linear stability of the rotating wave solution (\ref{eq:rotwave})
of the ring (\ref{eq:ring}) is defined by the variational equation
\begin{eqnarray*}
\frac{d\delta_{n}(t)}{dt} & = & \partial_{1}f(h(t+n\theta),h(t+(n-1)\theta-\sigma))\delta_{n}(t)+\\
 & + & \partial_{2}f(h(t+n\theta),h(t+(n-1)\theta-\sigma))\delta_{n-1}(t-\sigma),
\end{eqnarray*}
where $\delta_{n}(t)\in\mathbb{R}^{d}$ is the variation of $x_{n}$
Introducing the delay compatibility condition (\ref{eq:delaycomp})
in the form $\tau=\sigma+\theta$ into the coefficient matrices, and
substituting the Floquet ansatz $\delta_{n}(t)=r_{n}(t)e^{\lambda t}$,
we obtain 
\begin{eqnarray}
\frac{dr_{n}(t)}{dt} & = & \left(A(t+n\theta)-\lambda\mathrm{Id}\right)r_{n}(t)+\label{eq:drn}\\
 &  & +e^{-\lambda\sigma}B(t+n\theta)r_{n-1}(t-\sigma),
\end{eqnarray}
where $r_{n}(t)$ are $T$-periodic functions and $\lambda$ the characteristic
exponents to be found. A component-wise time-shift transformation
\[
p_{n}(t)=r_{n}(t-n\theta)
\]
(see \cite{Luecken2013b} for details) leads to 
\[
\frac{dp_{n}(t)}{dt}=(A(t)-\lambda\mathrm{Id})p_{n}(t)+e^{-\lambda\sigma}B(t)p_{n-1}(t-\tau).
\]
The coupling matrix of the unidirectional ring can be diagonalized
by passing to discrete Fourier modes 
\[
p_{n}(t)=\sum_{m=1}^{N}\hat{p}_{m}(t)e^{2\pi imn/N}
\]
such that we obtain the decoupled spectral problems 
\begin{equation}
\frac{d\hat{p}_{m}(t)}{dt}=(A(t)-\lambda_{m}\mathrm{Id})\hat{p}_{m}(t)+e^{-\lambda_{m}\sigma-i\psi_{m}}B(t)\hat{p}_{m}(t-\tau).\label{eq:eig2}
\end{equation}
for $m=1,...,N$, which are only distinguished by the phase factor
\[
\psi_{m}=2\pi m/N
\]
and have all a similar form as the spectral problem for the single
equation (\ref{eq:eig1}). As it is shown in Lemma 5 in \cite{Sieber2013},
the characteristic exponents of a Floquet problem (\ref{eq:eig1})
for the single equation (\ref{eq:1}) can be obtained from a characteristic
equation of the form 
\begin{equation}
F\left(\lambda,e^{-\lambda\tau_{k}}\right)=0.\label{eq:ch1}
\end{equation}
with some analytic function $F$. This function can be used also to
determine the limiting spectra for $k\rightarrow\infty$, which is
given by 
\begin{equation}
F\left(\lambda,0\right)=0.\label{eq:chstrong}
\end{equation}
for the strongly unstable spectrum and 
\begin{equation}
F\left(i\omega,e^{-\gamma-i\varphi}\right)=0\label{eq:F2}
\end{equation}
for the asymptotic continuous spectral curves approximating weak spectrum
with the limiting behavior 
\begin{equation}
\lambda=i\omega+\frac{\gamma}{\tau}\label{eq:spec_scal}
\end{equation}
for real $\omega$ and $\gamma$. These curves are parametrized by
the extra parameter $\varphi$ in (\ref{eq:F2}). For the decoupled
spectral problems (\ref{eq:eig2}) of the ring, we obtain the same
function $F(\cdot,\cdot)$ with the second argument replaced by $e^{-\lambda\sigma-i\psi_{m}}$.
However, the resulting equations (\ref{eq:chstrong}) and (\ref{eq:F2})
for the limiting spectra coincide, since the phase factor $\psi_{m}$
can be absorbed into the curve parameter $\varphi$. Hence, for large
delay the stability properties of the periodic solution in the single
equation (\ref{eq:eig1}) coincides with the stability of all rotating
waves independent on their wave number $M$.

\section{Instantaneous coupling in the ring of oscillators\label{sec:Instantaneous-coupling-in}}

So far, we studied the situation with a delay both in the single oscillator
and in the unidirectionally coupled ring. However, the delay compatibility
condition (\ref{eq:delaycomp}) can be satisfied also with $\sigma=0$,
i.e. instantaneous coupling in the ring. This happens exactly, if
for the single oscillator, we have a periodic orbit $h(t)$ with a
period $T$ resonant to the delay $\tau_{0}$ 
\begin{equation}
M_{0}T/N_{0}=\tau_{0}.\label{eq:res_delay}
\end{equation}
Assuming that $M_{0}$ and $N_{0}$ are relative prime, the periodic
solution can be embedded into a ring of minimal size $N_{0}$, and
additionally for all integer multiples 
\[
N_{j}=jN_{0}
\]
with the wave number being the corresponding multiple of $M_{0}$.

\begin{figure}
\includegraphics{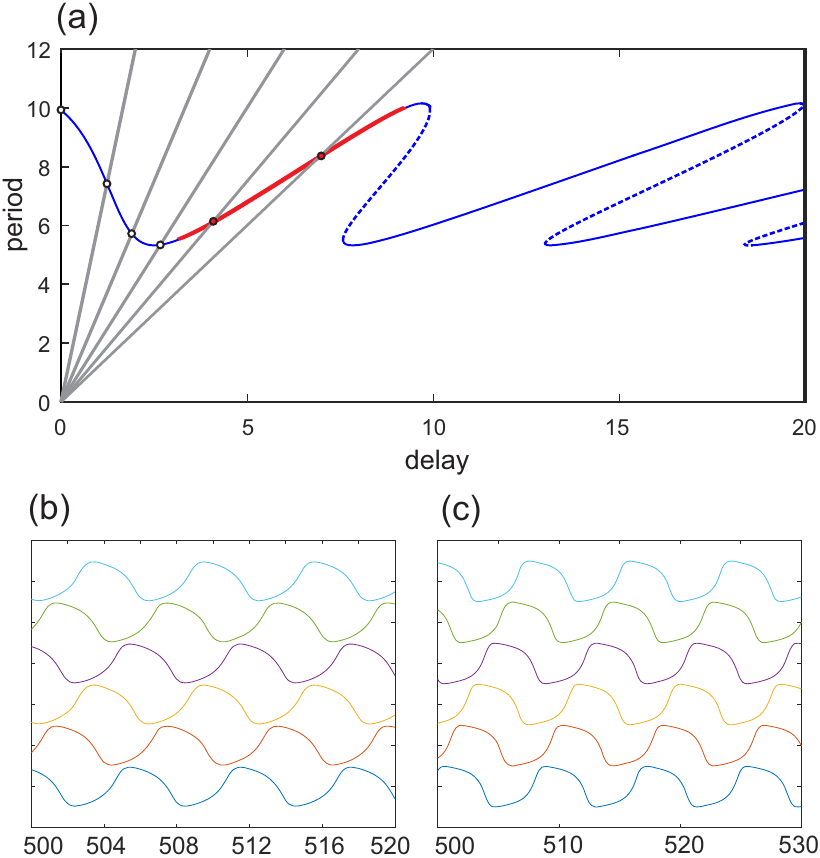}

\caption{\label{fig:vdp0_bd_lines}(a) Branch of periodic solutions for a single
Van-der-Pol oscillator with delayed feedback (\ref{eq:VdP}). Solid
and dashed parts indicate stability and instability, respectively.
The red thick part of the line corresponds to periodic solutions that
are stable when they reappear for sufficiently large delays. Grey
lines indicate the resonance conditions $\tau=MT/N$ for $N=6$ and
$M=1,\dots,5$ and their intersections with the branch indicates resonant
periodic solutions leading to rotating waves. 
Stable rotating waves exist for $M=4$ (b) and $M=5$ (c).}
\end{figure}

In Fig.~\ref{fig:vdp0_bd_lines}(a) we show that a branch of periodic
solutions with varying delay generically leads to the appearance of
resonances of the period and the delay. In the figure the resonance
conditions for $N=6$ are shown as gray lines. Their intersections
with the branch of periodic solutions for a single Van-der-Pol oscillator
with delayed feedback (\ref{eq:VdP}) correspond to resonant periodic
solutions. The thin solid line indicates the stable part of the branch,
while the red thick line corresponds to periodic solutions that are
stable when they reappear for sufficiently large delays. As we will
show below, this implies the stability of the corresponding rotating
waves. In this case, stable rotating waves exist only for $M=4$ and
$M=5$, shown in panels (b) and (c). Note that the rotating waves
with different wave numbers are unstable even although the resonant
periodic orbits for some of them lie on the stable part of the branch.

As we have seen, the stability of the rotating wave may be different
from the stability of the original periodic solution at $\tau_{0}$.
However, we will show now that the stability of the rotating wave
relates to the stability of $h(t)$ for large delays $\tau_{k}$:
\emph{If the resonant periodic solution $h(t)$ is asymptotically
stable for all large enough $\tau$, then the rotating wave in the
ring with no delays is stable as well.} 

To prove this, note that the stability of $h(t)$ for large delay
implies that the branches of asymptotic continuous spectrum (\ref{eq:spec_scal}),
given by (\ref{eq:F2}), have negative real parts except for a single
zero at $\omega=\varphi=0$ corresponding to the trivial Floquet-mode
given by the shift along the periodic orbit itself (see \cite{Sieber2013},
Lemma 18). Now, assume that the rotating wave is unstable, i.e. the
characteristic equation 
\[
F\left(\lambda,e^{-\lambda\sigma-i\psi_{m}}\right)=0,
\]
has a solution $\lambda$ with $\text{Re}\lambda>0$ for $\sigma=0$.
Then, by increasing $\sigma$, this solution must cross the imaginary
axis at some point $i\Omega$ for some value $\sigma$, since for
large $\sigma$ the rotating wave is stable as it is shown in Section~\ref{sec:stab_prop}.
According to (\ref{eq:ch1}) this solution satisfies 
\[
F\left(i\Omega,e^{-i\Omega\sigma-i\psi_{m}}\right)=0,
\]
for some wave number $m$ of the perturbation. This contradicts to
the assumption of stability of $h(t)$ for large delay, which implies
that $F\left(i\omega,e^{-i\varphi}\right)\neq0$ for all values of
$\omega$ and $\varphi$ different from zero. Thus, the asymptotic
stability for large delay of a periodic solution of the single oscillator
(\ref{eq:1}) guarantees the stability of the corresponding rotating
waves with instantaneous coupling in (\ref{eq:ring}). Note that this
is true for all integer multiples of the minimal ring size $N_{0}$.

Choosing a sufficiently large size $N_{j}$ of the ring, it is also
possible to show the opposite: \emph{If the resonant periodic solution
$h(t)$ of the single oscillator (\ref{eq:1}) is weakly unstable
for large delay $\tau_{k}$, then the corresponding rotating wave
is also unstable in a ring (\ref{eq:ring}) of sufficiently large
size $N_{j}$ and coupling delay $\sigma=0$.} Indeed, in the ring
with $\sigma=0$ the characteristic equation for the rotating waves
takes form 
\begin{equation}
F(\lambda,e^{-i\psi_{m}})=0,\label{eq:ch3}
\end{equation}
Stability of the rotating wave would imply that all roots of (\ref{eq:ch3})
except for $\psi_{0}=0$ have negative real part. Since for large
$N$ the phases $\psi_{m}$ densely fill the interval $[0;2\pi]$,
by analyticity of $F$ all roots of 
\[
F(\lambda,e^{-i\varphi})=0
\]
with $0<\varphi<2\pi$ have then also negative real part. At the other
hand, if the periodic solution $h(t)$ of (\ref{eq:1}) is weakly
unstable for long delay $\tau_{k}$, there must be some point $\Omega\neq0$
for which the pseudo-continuous spectrum crosses the imaginary axis,
i.e. $\gamma(\Omega)=0$ and hence $F(i\Omega,e^{-i\varphi})=0$.
This contradiction proves that the rotational wave is in fact unstable.

Thus, the stability of a periodic solution of the single oscillator
at large delay is a sufficient condition for the stability of the
corresponding rotating wave in a ring with instantaneous coupling.
For a large number $N$ of oscillators, it is also a necessary condition.

\section{Jittering regimes in a ring of oscillators with instantaneous coupling\label{sec:Jittering-regimes-in}}

\subsection{Theoretical study of the multi-jittering }

We have shown in the previous sections that periodic solutions of
a single oscillator with delayed feedback can be embedded as rotating
waves in a ring with instantaneous couplings. This fact implies that
regimes that have been observed in time-delayed systems, are generally
expected in complex networks, that do not have necessary time delays.
The reverse statement is also true: dynamical phenomena observed in
lattices of coupled oscillators are also expected in systems with
large delays, such as chimera states \cite{Larger2013}, reservoir
computing \cite{Larger2012}, Eckhaus instability \cite{Wolfrum2006},
complex pseudo-spatial patterns \cite{Giacomelli1996}, etc.

In this section, we concentrate on a fascinating phenomena that recently
was discovered in oscillatory systems with long delays, the so-called
multi-jittering instability \cite{Klinshov2015PRL}. Later similar
phenomena were observed in rings with time delays \cite{Klinshov2016}.
Now we show that the multi-jittering can be observed in a ring with
instantaneous couplings.

\begin{figure}
\begin{centering}
\includegraphics{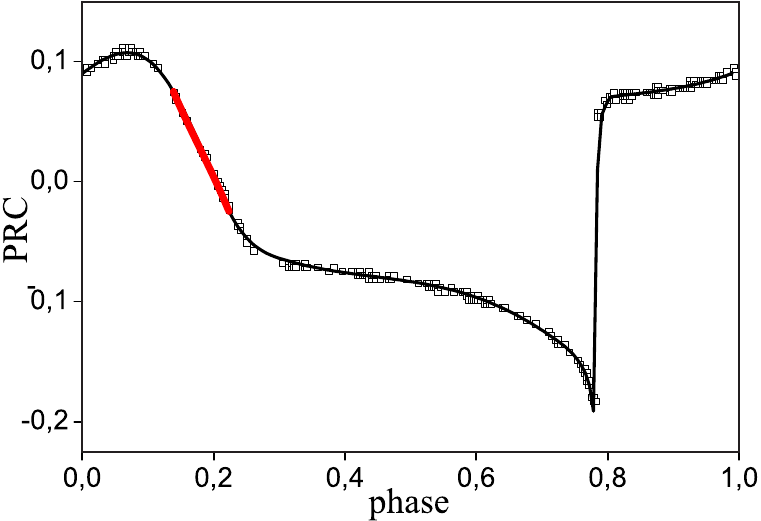}
\par\end{centering}
\caption{\label{fig:prc}The PRC of the electronic FitzHugh-Nagumo oscillator.}
\end{figure}

The multi-jittering was first discovered in a single oscillator with
delayed pulse feedback. The minimal model is a phase oscillator, where
the action of the pulse feedback is modeled with the so-called phase
resetting curve (PRC) $Z(\varphi)$, which characterizes the response
of the oscillator to an incoming pulse at different phases:
\begin{equation}
\frac{d\varphi}{dt}=1+Z(\varphi)\sum_{t_{p}}\delta\left(t-t_{p}-\tau\right).\label{eq:jit1}
\end{equation}
Here, $\varphi$ is the oscillator's phase, $Z(\varphi)$ its PRC,
and $\tau$ the delay. In the absence of the feedback, the phase grows
uniformly with $d\varphi/dt=1$. When the phase reaches unity, it
resets to zero, and the oscillator emits a spike. The moments of spike
emissions are denoted as $t_{p}$. Each spike is sent to a delay line,
and it arrives to the oscillator after the delay $\tau$ causing an
instantaneous phase shift $\text{\ensuremath{\Delta\varphi}=}Z(\varphi)$.
Further, we select the PRC $Z(\varphi)$ as in Fig.~\ref{fig:prc}
which corresponds to our setup of the electronic FitzHugh-Nagumo oscillator
\cite{Klinshov2015PRL,Klinshov2016}. 

\begin{figure}
\begin{centering}
\includegraphics{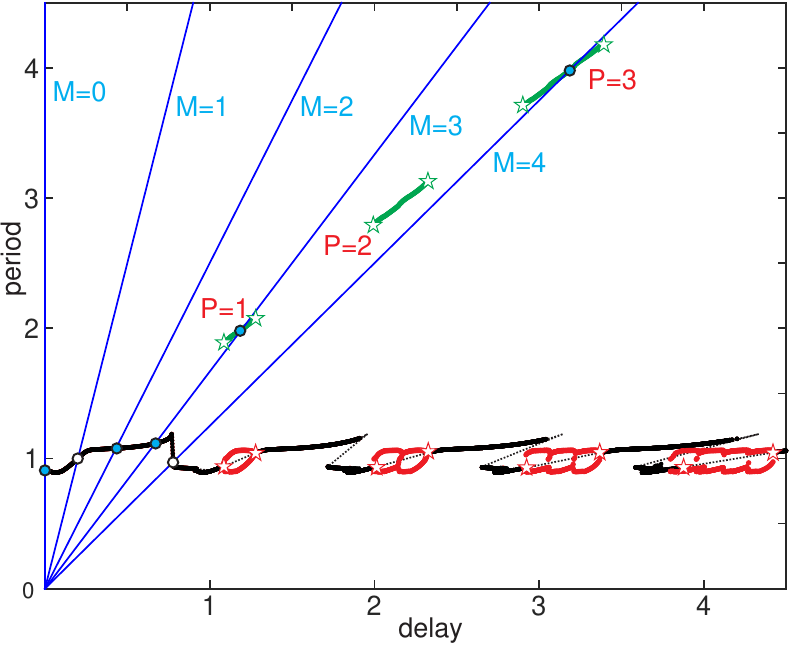}
\par\end{centering}
\caption{\label{fig:jit1}The bifurcation diagram of a single oscillator with
pulse delayed feedback (\ref{eq:jit1}), period versus the delay.
Black solid line corresponds to stable regular regimes, gray dashed
line to unstable regular regime, green lines to jittering regimes.
Red solid lines depict the inter-spike intervals that constitute the
jittering regimes. Stars denote the multi-jitter bifurcation points.
The bifurcation diagram is combined with a family of lines $\tau=MT/N$
with $N=5$. The intersection points correspond to rotating waves
in a ring of pulse-coupled oscillators with no delays (\ref{eq:jit_ring0}).
Hallow points correspond to unstable, solid to stable waves.}
\end{figure}

The bifurcation diagram for (\ref{eq:jit1}) is shown in Fig.~\ref{fig:jit1}
where the periods of the observed regimes are plotted versus the delay.
The most common regime is the so-called regular spiking when the oscillator
emits spikes with constant inter-spike interval. The stable parts
of the corresponding branch are plotted by black solid line, unstable
by gray dashed line. 

In \cite{Klinshov2015PRL} we have shown that under quite general
conditions the regular spiking regime may be destabilized via a very
peculiar scenario. Namely, if the PRC function $Z(\varphi)$ has a
point $\psi^{*}$ with slope $Z'\left(\psi^{*}\right)=-1$, the so-called
``multi-jitter'' bifurcations take place for delay values 
\begin{equation}
\tau_{J}=P\left(1-Z\left(\psi^{*}\right)\right)+\psi^{*},\label{eq:jitbif}
\end{equation}
where $P\in\mathbb{N}$. In each of these bifurcations, the regular
spiking solution destabilizes, and the so-called ``jittering'' solutions
emerge. These regimes are characterized by a periodic sequence of
non-equal inter-spike intervals. It was shown that the length of one
period in the sequences equals $P+1$ inter-spike intervals, i.e.
grows linearly with the delay (compare to (\ref{eq:jitbif})). In
Fig.~\ref{fig:jit1} the periods of the jittering solutions are plotted
by green lines, while the inter-spike intervals constituting them
are plotted by red. The latter curves branch off the regular spiking
branch at the multi-jitter bifurcation points denoted by stars.

In \cite{Klinshov2016}, we described the jittering instability in
rings of pulse delay coupled oscillators. It was shown that rotating
waves in such rings may destabilize and give birth to ``jittering''
waves. Here we will demonstrate that the jittering waves may appear
in rings without delays.

Consider a ring of oscillators with unidirectional instantaneous pulse
coupling
\begin{equation}
\frac{d\varphi_{n}}{dt}=1+Z(\varphi_{n})\sum_{t_{n-1,p}}\delta\left(t-t_{n-1,p}\right),\label{eq:jit_ring0}
\end{equation}
where $n=1,...,N.$ Here, the spikes produced by $(n-1)$-st oscillator
at moments $t_{n-1,p}$ immediately arrive to the $n$-th oscillator
and perturb it.

\begin{figure}
\begin{centering}
\includegraphics{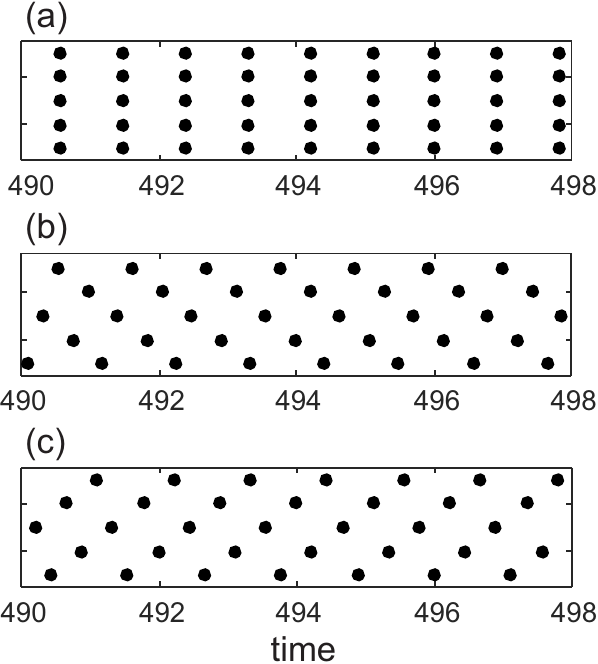}
\par\end{centering}
\caption{\label{fig:regwaves}Regular rotating waves in (\ref{eq:jit_ring0}):
(a) $M=0$, (b) $M=2$ and (c) $M=3$. Points denote the moments of
spikes of the oscillators.}
\end{figure}

The ring (\ref{eq:jit_ring0}) consists of the same oscillators as
(\ref{eq:jit1}), thus the periodic solutions of (\ref{eq:jit1})
transfer into rotating waves of (\ref{eq:jit_ring0}). To obtain the
rotating wave solutions, the bifurcation diagram in Fig. \ref{fig:jit1}
is complemented by the family of lines $\tau=MT/N$ (similarly to
Fig.~\ref{fig:vdp0_bd_lines}), where $M=0,1,...N-1$, and $N=5$.
The points where these lines intersect the branches of the periodic
solutions correspond to the rotating waves. As one can see in Fig.~\ref{fig:jit1},
there are 5 points where the regular spiking branch of $T(\tau$)
intersects one of the lines $\tau=MT/N$. We have checked that some
of these points correspond to stable rotating waves in the ring. These
points are marked by blue circles and appear on the asymptotically
stable part of the branch. The other points correspond to unstable
rotating waves and are marked by white circles. The stable regular
waves are observed in the system for $M=0,2,3$, they are illustrated
in Fig.~\ref{fig:regwaves}(a-c).

\begin{figure}
\begin{centering}
\includegraphics{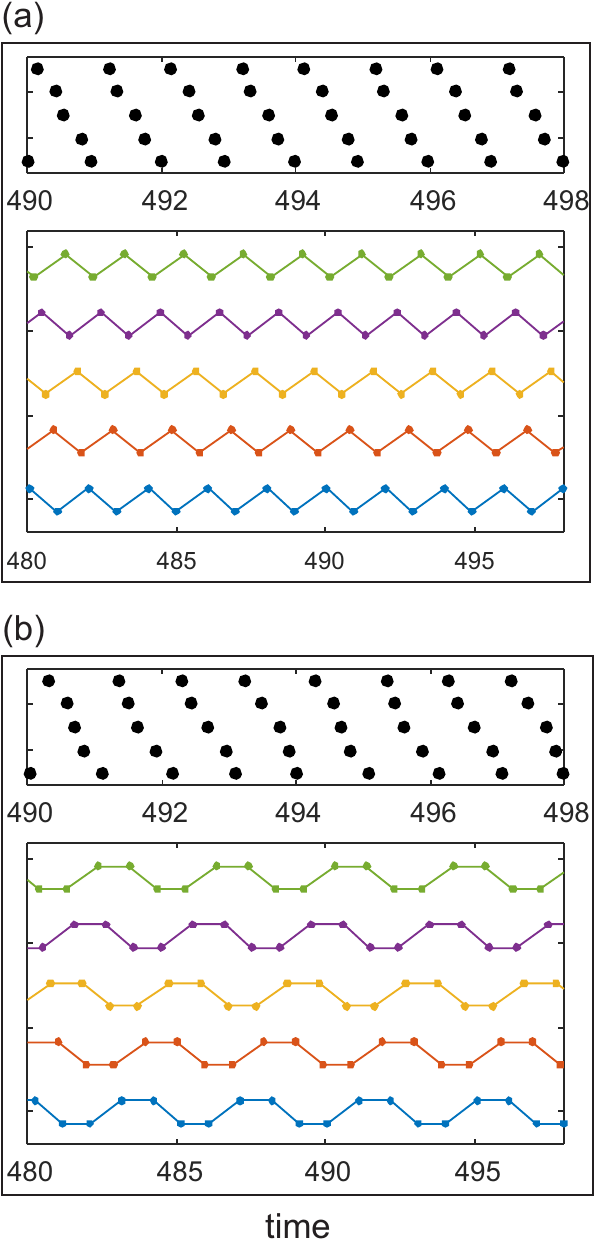}
\par\end{centering}
\caption{\label{fig:jitwaves}Two different jittering rotating waves observed
in (\ref{eq:jit_ring0}). The top panels depict the moments of spike
generation, the bottom inter-spike intervals.}
\end{figure}

The lines in Fig.~\ref{fig:jit1} also intersect the branches of
jittering solutions which implies the existence of jittering rotating
waves. These waves have indeed been observed, they are illustrated
in Fig.~\ref{fig:jitwaves}(a,b). In these Figures, the top panels
show the moments of spikes emission by the oscillators, while the
bottom panels depict the dynamics of the inter-spike intervals. The
sequences of inter-spike intervals are the same for all the oscillators,
although time-shifted, and consist of two distinct values (the so-called
bipartite solutions). Note that the plots corresponding to the different
oscillators are shifted along the vertical axis for the sake of better
representation. In reality the values of the short and the long inter-spike
intervals are also the same for all the oscillators. Thus, the rotating
wave of inter-spike intervals propagates in the ring.

\subsection{Experimental demonstration in a ring of electronic FitzHugh-Nagumo
oscillators\label{subsec:Experimental-demonstration-in}}

We have also corroborated our theoretical results by the experiments
with electronic circuits. We have experimentally studied a ring of
$N=5$ electronic FitzHugh-Nagumo oscillators with instantaneous unidirectional
couplings. The circuitry of the single electronic oscillator used
in the experiment is the same as \cite{Klinshov2014}, while the parameters
of the oscillators are set as in \cite{Klinshov2015PRL}. In the absence
of coupling each oscillator emits spikes periodically with the period
$T\approx2.95$ ms (the frequency mismatch between the oscillators
is about 0.4\%). The coupling in the ring is organized as follows:
when the output voltage of each oscillator exceeds the threshold value
$u_{th}=0.84$V, a square-shape pulse of the amplitude $A_{P}=5$V
and duration $T_{P}=42\mu$s is sent to the next oscillator. The phase
resetting curve corresponding to such the pulse is depicted in Fig.
\ref{fig:prc}. 

\begin{figure}
\begin{centering}
\includegraphics{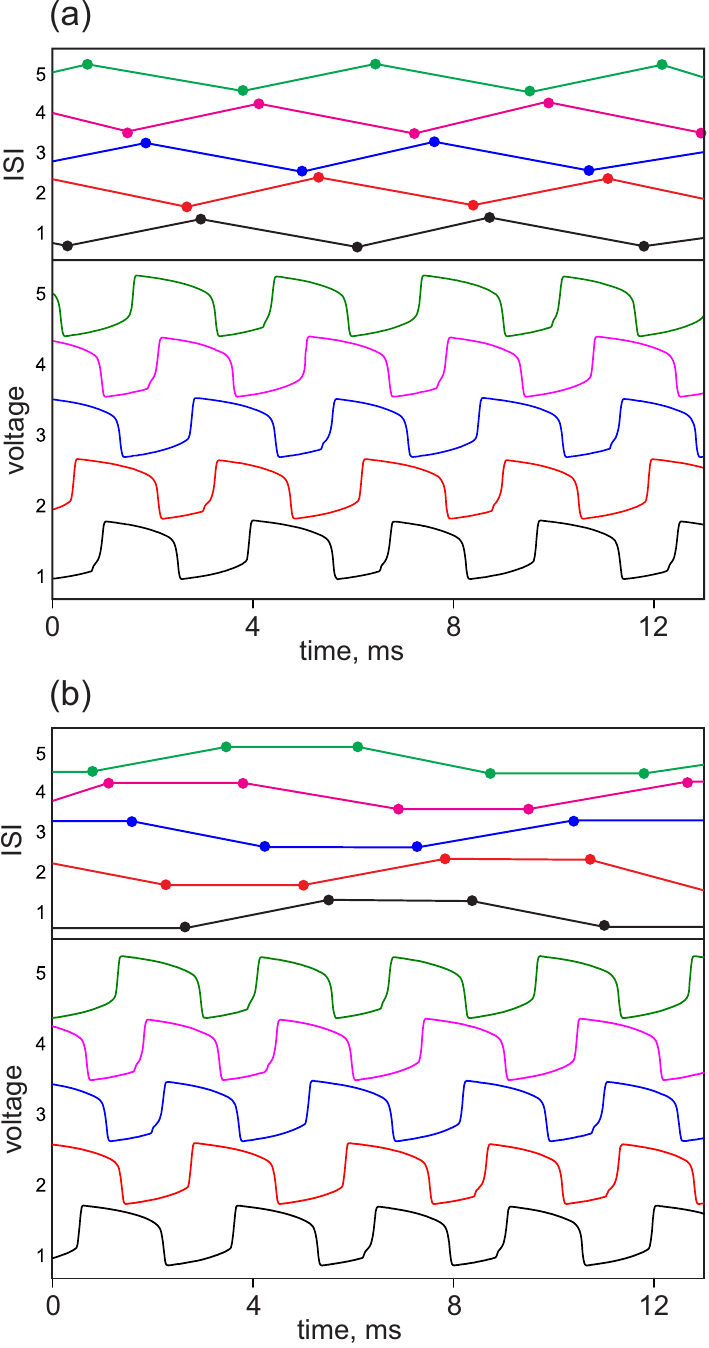}
\par\end{centering}
\caption{\label{fig:jitwaves-exp}Two different jittering rotating waves experimentally
observed in a ring of five electronic FitzHugh-Nagumo oscillators.
The top panels depict the inter-spike intervals, bottom ones the output
voltages of the oscillators.}
\end{figure}

In the experiment, we started from random initial conditions and traced
the dynamics of the output voltages and the inter-spike intervals
of all the oscillators. We were able to detect all the regimes predicted
theoretically, except for the global synchrony (Fig.~\ref{fig:jitwaves-exp}(a)).
The presumable reason of not finding this regime is its low stability
and high sensitivity to the frequency detuning. In Fig.~\ref{fig:jitwaves-exp}
the examples of two distinct jittering rotating waves are given.

We have demonstrated jittering rotating waves in the ring of pulse-coupled
oscillators with no coupling delays. Previously, we have shown that
jittering solutions emerge in the oscillator with delayed feedback
generically, and the only necessary condition for that is the presence
of parts with slope $<-1$ in the oscillator's PRC. In this case the
appropriate choice of the delay allows to establish a jittering solution
of arbitrary long period. Further we will show that jittering waves
generically appear in rings of oscillators with instantaneous pulse
coupling.

\subsection{Generic appearance of jittering waves in rings without delays}

In this section we show that under the assumption that one delayed
pulse coupled oscillator (\ref{eq:jit1}) has a jittering regime,
it is always possible to select a large enough number $N$ of oscillators
in a ring (\ref{eq:jit_ring0}) without delays such that this ring
possesses a jittering wave.

As shown in \cite{Klinshov2015PRL}, the branch of regular spiking
solutions can be determined parametrically as $T(\psi)$, $\tau(\psi)$,
where 
\begin{eqnarray*}
T(\psi) & = & 1-Z(\psi),\\
\tau(\psi) & = & P\left(1-Z(\psi)\right)+\psi,
\end{eqnarray*}
 $P\in Z$, $T(\psi)$ is the inter-spike interval at time-delay $\tau(\psi)$
The stability of these regular spiking solutions is given by the condition
$-1<\alpha<1/P$, where $\alpha:=Z'(\psi)$. For $\alpha=1/P$, the
regular solutions undergo a fold bifurcation, and for $\alpha=-1$
the multi-jitter bifurcation. Jittering solutions branch from the
regular ones at the multi-jitter bifurcation points, i.e. at $\tau(\psi^{*})=P\left(1-Z\left(\psi^{*}\right)\right)+\psi^{*}$,
where $Z'\left(\psi^{*}\right)=-1$. Each jittering regime consists
of distinct inter-spike intervals $\Theta_{j}$ that form a periodic
sequence. This sequence has a length $P+1$, thus the period of the
jittering solution equals $T=\Theta_{1}+\Theta_{2}+...+\Theta_{P+1}$.
At the bifurcation point, all the inter-spike intervals tend to the
period of the regular solution from which they have branched. Since
the period of the regular solution equals $T_{R}=1-Z\left(\psi^{*}\right)$,
the period of the emergent jittering solution at the bifurcation point
equals $T_{J}=(P+1)\left(1-Z\left(\psi^{*}\right)\right)$.

Easy to show that at the bifurcation point the period of the emergent
jittering solution $T$ is larger than the delay $\tau$. Each branch
of jittering solutions connect two points of multi-jitter bifurcations
A and B. Assume without loss of generality that $\tau_{A}/T_{A}<\tau_{B}/T_{B}<1$.
Then it is always possible to select a rational number $M/N\in I=[\tau_{A}/T_{A},\tau_{B}/T_{B}]$
so that the line $\tau/T=M/N$ intersects the branch. This would assure
that the corresponding jittering solution can be embedded as a jittering
rotating wave in the ring of $N$ oscillators with wave number $M$
(see (\ref{eq:res_delay})).

Thus, we have shown that jittering rotating waves generically emerge
in a rings with instantaneous pulse coupling. It is always possible
to take a sufficiently large number of oscillators and obtain a wave
with an arbitrary large period. Let us estimate the minimal size $N$
of the ring which is necessary to embed the jittering solution. Substituting
the expressions for the bifurcation parameters $\tau_{A,B}$ and $T_{A,B}$
one obtains

\[
I=\left[1-\frac{f\left(\psi_{A}\right)}{(P+1)},1-\frac{f\left(\psi_{B}\right)}{(P+1)}\right],
\]
where $f(\varphi)=1-\varphi/\left(1-Z(\varphi)\right)$. As $P$ grows,
the interval $I$ shrinks and approaches unity. From this observation
it is easy to see that the minimal size $N$ is proportional to the
jittering period $P$. Moreover, for large $P$ the wave number $M$
must be close to $N$.

\section{Conclusions }

In this paper, we have studied the relation between the dynamics of
a single oscillator with delayed feedback and a feed-forward ring
of identical oscillators. The obtained results may be summarized as
follows:

1. Periodic solutions of the single oscillator appear as rotating
waves in the ring with coupling delays $\sigma=\tau-MT/N+kT$. Here,
$k\in\mathbb{Z}$ and $M=0,...,N-1$ is the wave-number of the rotating
wave.

2. In particular, when the period is rationally related to the delay
such as $\tau=MT/N$, the periodic solution embeds as a wave in the
ring with instantaneous coupling $\sigma=0$.

3. The stability of the rotation waves in the ring for large delays
$\sigma$ is equal to the stability of the corresponding periodic
solution in the single oscillator for large delays $\tau.$

4. For large number $N$ of units in the ring, the stability of the
rotation wave for $\sigma=0$ coincides with the stability of the
corresponding periodic solution for $\tau$$\to\infty$. For small
$N$, stability of the periodic solution for $\tau\to\infty$ is a
necessary condition for the stability of the wave for $\sigma=0$. 

The discovered relations between the dynamics of a single oscillator
with delayed feedback and a ring of oscillators are of great interest.
Particularly, it suggests that dynamical regimes observed in systems
with delays may be also observed in systems without delays but with
a ring topology. As an example we considered high-periodical jittering
regimes in oscillators with pulse delayed feedback. We show that these
regimes generically can be embedded as jittering rotating waves in
rings of pulse-coupled oscillators without coupling delays.
\begin{acknowledgments}
V.K., D.S., S.Y. and V.N. acknowledge the support from the Russian
Scientific Foundation (project 16-42-01043 for the Institute of Applied
Physics) and the German Research Foundation (project SCHO 307/15-1
and YA 225/3-1 for TU Berlin).
\end{acknowledgments}

\end{document}